# Thermodynamic properties of ε-Fe with thermal electronic excitation effects on vibrational spectra

Jingyi Zhuang[1,2], Hongjin Wang[3,4], Qi Zhang[4], Renata M. Wentzcovitch[1,2,4]

[1] *Department of Earth and Environmental Sciences, Columbia University, New York, NY, 10027, USA*

[2] *Lamont–Doherty Earth Observatory, Columbia University, Palisades, NY, 10964, USA*

[3] *Department of Computer Sciences, Columbia University, New York, New York, USA, 10027*

[4] *Department of Applied Physics and Applied Mathematics, Columbia University, New York, New York, USA, 10027*

**Abstract**

The thermodynamic properties of hcp-iron (ε-Fe) are essential for investigating planetary cores' internal structure and dynamic properties. Despite their importance to planetary sciences, experimental investigations of ε-Fe at relevant conditions are still challenging. Therefore, *ab initio* calculations are crucial to elucidating the thermodynamic properties of this system. Here we use a free energy calculation scheme based on the phonon gas model compatible with temperature-dependent phonon frequencies. We investigate the effects of electronic thermal excitations, which introduces a temperature dependence on phonon frequencies, and the implication for the thermodynamic properties of ε-Fe at extreme pressure ($P$) and temperature ($T$) conditions. We disregard phonon-phonon interactions, i.e., anharmonicity and their effect on phonon frequencies. Nevertheless, the current scheme is also applicable to $T$-dependent anharmonic frequencies. We conclude that the impact of thermal electronic excitations on vibrational properties is not significant up to ~ 4,000 K at 200 GPa but should not be ignored at higher temperatures or pressures. However, the static free energy, $F_{st}$, must always include the effect of thermal excitation fully in a continuum of $T$s. Our results for isentropic equations of state show good agreement with data from recent ramp compression experiments up to 1,400 GPa conducted at the National Ignition Facility (NIF).

# I. INTRODUCTION

HCP-iron (ε-Fe) is the likely stable phase of iron at the extreme conditions of terrestrial solar and extra-solar planetary cores [1]. Its thermodynamic properties are essential for modeling this region and the mantle above it. There are numerous pioneering theoretical and experimental studies of the properties of ε-Fe. Previous *ab initio* calculations of ε-Fe have been carried out using various techniques, e.g., Quantum Monte Carlo (QMC) [2], all-electron methods [3], pseudopotential methods [4–9], molecular dynamics (MD) [10], etc. Experimental investigations of equations of state (EoS) of ε-Fe include static compression in diamond anvil cell (DAC) [11–13], and dynamic compression experiments [14,15] up to 300 GPa. A recent ramp compression experiment [16] measured the density-pressure relation in ε-Fe up to 1.4 TPa at unconstrained temperatures. Planetary cores reach very high temperatures, e.g., ~6,000 K at ~365 GPa in the Earth [17], and isothermal compression results are also significant. Recent modeling of terrestrial exoplanets with up to 20 Earth masses [18] shows that pressures and temperatures at the center of these planets and newly identified phases can reach over 13 TPa and 9,000 K. These conditions remain challenging also for computations of solid-state properties as such high temperatures bring extra complexity, e.g., anharmonicity, electronic thermal excitations, and likely atomic diffusion [19]. In previous *ab initio* studies [8,20–22], some of these effects were not fully considered, or their implications were not analyzed. Here we investigate the influence of electronic excitations alone on the vibrational and thermodynamic properties of ε-Fe. Similar to phonon-phonon interaction, it produces temperature-dependent (*T*-dependent) phonon frequencies that also affect free energy calculations.

Thermodynamic properties of ε-Fe addressing these complex effects have been discussed in some previous *ab initio* studies. Unavoidably, results differ somewhat because of multiple methodologies used [5,6,22]. Among the popular methods, the quasiharmonic approximation (QHA) is appropriate for addressing these properties in weakly anharmonic solids up to ~ 2/3 of the melting temperature in most cases. It is computationally less demanding than MD. It requires only calculations of the vibrational density of states (VDoS) at about ten pressures. Such calculations are more challenging for metals where electronic thermal excitations may affect phonon frequencies [23].

Hence, we implemented a free energy calculation scheme based on the phonon gas model (PGM) compatible with *T*-dependent phonon or phonon quasiparticle frequencies to investigate ε-Fe at planetary interior conditions. The present calculation disregards phonon-phonon interaction effects, i.e., anharmonicity, but addresses directly and precisely the unavoidable impact of electronic thermal excitations on phonon-dispersions and thermodynamic properties of metals. Nonetheless, the present scheme is also applicable to free energy computations when phonon-phonon interactions are non-negligible and the *T*-dependence of phonon quasiparticle frequencies originates in anharmonicity [21,24]. The current implementation offers properties in a continuum range of states up to ultra-high temperatures and pressures [24]. Here we present thermodynamic properties of ε-Fe covering a wide range of pressures (0 - 1,400 GPa) and temperatures (0 - 8,000 K). In particular, we investigate the interplay between electronic thermal excitations and vibrational properties and the effect it plays on thermodynamic properties. Results for the isentropic EoS of ε-Fe are in good agreement with data from recent ramp compression experiments up to 1,400 GPa conducted at the National Ignition Facility (NIF) [16].

This paper is organized as follows. In the next section, we introduce the *ab initio* simulation details and the phonon gas model formalism. The following section shows results and compares them with experimental data and previous calculations. We summarize our conclusions in the last section.

## II. METHODS

### 1. Phonon gas model (PGM)

The PGM assumes phonons or phonon quasiparticles do not interact. Within the Born-Oppenheimer approximation and for harmonic systems, the PGM free energy is given by the quasiharmonic approximation (QHA):

$$F(V,T) = F_{st}(V) + F_{vib}(V,T), \quad (1a)$$

For insulators

$$F_{st}(V) = E_{KS}(V), \quad (1b)$$

where $F_{st}(V)$ is the static free-energy, with $E_{KS}(V)$ being the Kohn-Sham energy for an equilibrium ionic configuration with static equilibrium volume *V*.

$$F_{vib}(V,T) = \frac{1}{2}\sum_{q,s} \hbar\omega_{q,s}(V) + k_B T \sum_{q,s} \ln\left(1 - \exp\left(-\frac{\hbar\omega_{q,s}(V)}{k_B T}\right)\right), \quad (1c)$$

is the vibrational free-energy with $\omega_{q,s}(V)$ being the vibrational frequency of non-interacting phonons with wavenumber **q** and polarization index $s$, and $T = T_{ion}$, the ionic temperature. The 1st and 2nd terms on the r.h.s. of Eq. (1c) are the zero-point, $E_{zp}(V)$, and thermal vibrational free energy, $E_{th}(V,T)$, respectively. In this approximation, $\omega_{q,s}(V)$ is $T$-independent, depending only on $V$. The corresponding entropy of this system is:

$$S_{vib}(V,T) = k_B \sum_{q,s} \left( (1+n_{qs})\ln(1+n_{qs}) - n_{qs}\ln n_{qs} \right), \tag{1d}$$

where the normal mode population, $n_{qs}(V,T)$, is:

$$n_{qs}(V,T) = \frac{1}{\exp\frac{\hbar\omega_{qs}(V)}{k_B T} - 1}. \tag{1e}$$

For metallic systems with negligible anharmonicity (phonon-phonon interactions), the static energy, $F_{st}$, and vibrational frequencies should be computed with the Mermin functional [25,26], i.e., the finite temperature version of density functional theory (DFT). In this case, phonon frequencies acquire a $T$-dependence through thermal electronic excitation at a temperature $T_{el}$ [23]. For clarity's sake, we now distinguish $T_{el}$ and $T_{ion}$ to identify the origin of the $T$-dependence in the calculation. Obviously, $T_{el} = T_{ion} = T$, when the system is in thermodynamic equilibrium. A common, although approximate, expression for the free energy in this case,

$$F(V, T_{el}, T_{ion}) = F_{st}(V, T_{el}) + F_{vib}(V, T_{el}, T_{ion}), \tag{2a}$$

where

$$F_{st}(V, T_{el}) = F_{Mermin}(V, T_{el}), \tag{2b}$$

is the total Mermin free-energy for an equilibrium ionic configuration at volume $V$.

$$F_{Mermin}(V, T_{el}) = E_{st}(V, T_{el}) - T_{el} S_{el}(V, T_{el}), \tag{2c}$$

where $E_{st}(V, T_{el})$ is the self-consistent energy with orbital occupancies

$$f_{ki}(V, T_{el}) = \frac{1}{\exp\frac{\hbar(E_{ki} - E_F)}{k_B T_{el}} + 1}, \tag{2d}$$

with $E_{ki}$ being the one-electron energy of an orbital with wavenumber **k** and band index and $i$, and $E_F$ being the Fermi energy. The electronic entropy is

$$S_{el}(V,T_{el}) = -k_B \sum_{k,i}\big((1-f_{ki})\ln(1-f_{ki}) + f_{ki}\ln f_{ki}\big). \qquad (2e)$$

The vibrational energy is:

$$F_{vib}(V,T_{el},T_{ion}) = \tfrac{1}{2}\sum_{q,s}\hbar\omega_{q,s}(V,T_{el}=0) + k_B T_{ion}\sum_{q,s}\ln\left(1-\exp\left(-\tfrac{\hbar\omega_{q,s}(V,T_{el})}{k_B T_{ion}}\right)\right), \qquad (2f)$$

and the vibrational entropy is:

$$S_{vib}(V,T_{el},T_{ion}) = -\left.\frac{\partial F(V,T_{el},T_{ion})}{\partial T_{ion}}\right|_{T_{el},V} - \left.\frac{\partial F(V,T_{el},T_{ion})}{\partial T_{el}}\right|_{T_{ion},V}$$

$$= k_B \sum_{q,s}\big((1+n_{qs})\ln(1+n_{qs}) - n_{qs}\ln n_{qs}\big) - \sum_{q,s} n_{qs}\,\hbar\omega'_{q,s}(V,T_{el}), \qquad (2g)$$

with $\omega'_{q,s}(V,T_{el}) = \left.\frac{\partial \omega_{qs}(V,T_{el})}{\partial T_{el}}\right|_{T_{ion},V}$ and

$$n_{qs}(V,T_{el},T_{ion}) = \frac{1}{\exp\frac{\hbar\omega_{qs}(V,T_{el})}{k_B T_{ion}}-1}. \qquad (2h)$$

This type of calculation has previously been carried out for various systems, e.g., Fe [8] and Pt [27]. While $F_{st}(V,T_{el})$ has always been computed in a continuum $T_{el}$ range, $\omega_{qs}(V,T_{el})$ has often been computed at a single $T_{el}$, e.g., 300 K. As will be illustrated below, this approximation is satisfactory for nearly harmonic systems without rapidly varying electronic density of states at the Fermi level. In this case, $\omega'_{q,s}(V,T_{el})$ in Eq. (2g) may be negligible. While this procedure is a natural extension of the QHA to metallic systems, it is not correct because the vibrational entropy in Eq. (2g) differs from that in Eq. (1d), and the second term on the r.h.s. of Eq (2g) might be non-negligible. Eq. (1d) should always hold whether the system has $T$-independent or $T$-dependent frequencies [28,29].

For anharmonic insulators, the $T$-dependence of phonon frequencies originates in phonon-phonon interactions, i.e., anharmonicity. In this case, one replaces interacting phonons with non-interacting phonon quasiparticles with renormalized $T$-dependent frequencies $\tilde{\omega}_{qs}(V,T_{ion})$ and lifetimes $\tau_{qs}(V,T_{ion})$. The expressions for the particle population in Eq. (1e) and entropy in Eq. (1d) are still valid in this case [23,28,29]. Therefore, the entropy calculation should precede the free energy calculation. The free energy is obtained by integrating the entropy in this case. For an insulating system:

$$F(V,T_{ion}) = F_{st}(V,T_{ion}=0) + F_{vib}(V,T_{ion}), \qquad (3a)$$

where $F_{st}(V, T_{ion} = 0)$ is the static free-energy at $T = 0$ and

$$F_{vib}(V, T_{ion}) = F_{zp}(V, T_{ion} = 0) - \int_0^{T_{ion}} S_{vib}(V, T')dT', \tag{3b}$$

with $S_{vib}(V, T_{ion})$ given by Eq. (1d) with $T$-dependent frequencies, i.e.,

$$n_{\mathbf{q}s}(V, T_{ion}) = \frac{1}{\exp\frac{\hbar\widetilde{\omega}_{\mathbf{q}s}(V,T_{ion})}{k_B T_{ion}} - 1}. \tag{3c}$$

For metallic anharmonic systems,

$$F(V, T_{el}, T_{ion}) = F_{st}(V, T_{el}) + F_{vib}(V, T_{el}, T_{ion}), \tag{4a}$$

with $F_{st}(V, T_{el})$ given by Eq. (2c) and

$$F_{vib}(V, T_{el}, T_{ion}) = F_{zp}(V, T_{el} = 0, T_{ion} = 0) - \int_0^{T_{ion}} S_{vib}(V, T_{el} = T', T_{ion} = T')dT', \tag{4b}$$

where

$$S_{vib}(V, T_{el}, T_{ion}) = k_B \sum_{\mathbf{q},s} \left((1 + n_{\mathbf{q}s})\ln(1 + n_{\mathbf{q}s}) - n_{\mathbf{q}s}\ln n_{\mathbf{q}s}\right), \tag{4c}$$

with $n_{\mathbf{q}s}$ given by

$$n_{\mathbf{q}s}(V, T_{ion}, T_{el}) = \frac{1}{\exp\frac{\hbar\widetilde{\omega}_{\mathbf{q}s}(V,T_{ion},T_{el})}{k_B T_{ion}} - 1}. \tag{4d}$$

The total entropy of the metallic anharmonic system then is:

$$S_{tot}(V, T_{el}, T_{ion}) = S_{el}(V, T_{el}) + S_{vib}(V, T_{el}, T_{ion}). \tag{4e}$$

where $S_{el}(V, T_{el})$ given by Eq. (2e) is included in $F_{st}(V, T_{el})$ in Eq. (4a).

In this work, we compute the free energy of $\varepsilon$-Fe using Eqs. (2) and Eqs. (4), and compare their results. We refer to the scheme in Eqs. (4) as the entropy integration method (EIM) and to the scheme in Eqs. (2) as the $T$-dependent QHA (TQHA). We show that the $T_{el}$-dependence of the frequencies is weak, rendering the 2nd term on the r.h.s. of Eq. (2g) negligible, which makes both procedures practically equivalent. When this is the case, it is acceptable to neglect altogether the $T_{el}$–dependence of phonon frequencies as done in the past (e.g., Ref. [8]). However, the $T_{el}$-dependence of the static free energy, $F_{st}(V, T_{el})$, is crucial for obtaining accurate thermodynamic properties, especially the thermal expansivity [30]. We refer to this scheme as the temperature independent phonon (TIP). We also investigate a fourth scheme in which free energies and thermodynamic properties are obtained with constant $T_{el} = T$. Neither $F_{st}(V, T_{el})$ or phonon frequencies are $T_{el}$-dependent in this case. We refer to this scheme simply as QHA. These four schemes are summarized in Table I.

### 2. Simulation details

All calculations are performed using the Mermin functional [23,25,26] as implemented in the Quantum ESPRESSO [31] software. We use several available pseudopotential and PAW datasets [32–34], including the Evolutionary PAW (EPAW) dataset [32] as our primary choice, to investigate the isentropic EoS of ε-Fe up to 1,400 GPa and 8,000 K. As we focus on high pressure and temperature conditions, we conduct spin-unpolarized calculations. We use the Perdew-Burke-Ernzerhof generalized-gradient-approximation (PBE-GGA) [35] for the exchange and correlation functional. We choose a $10 \times 10 \times 10$ Monkhorst-Pack [36] **k**-point mesh and cutoff energy of 1,200 eV. Equilibrium structures at several electronic temperatures, $T_{el}$, and several static pressures are optimized using the damped variable cell-shape molecular dynamics [37,38] method. We choose $T_{el}$'s equal to 0 K, 300 K, and from 1,000 K to 8,000 K in steps of 1,000 K. For each $T_{el}$, structures are optimized and the vibrational densities of states (VDOS) computed at the following pressures: between 0 GPa and 300 GPa in steps of 50 GPa, between 300 GPa and 400 GPa in steps of 20 GPa, and between 500 and 1500 GPa in steps of 100 GPa. We calculate phonons on a $4 \times 4 \times 4$ **q**-point mesh, using density functional perturbation theory (DFPT) [39]. Brillouin Zone integration for entropy and free energy calculations is performed over a $10 \times 10 \times 10$ **q**-point mesh.

When using the EIM scheme expressed in Eqs. (4a-e), for each $V$ and $T_{el}$, the vibrational entropy is calculated for $T_{ion}$ varying from 0 K to 8,000 K in steps of 10 K. This entropy is then interpolated for $T_{el}$ on the same $T_{ion}$ grid using a spline interpolation. At each volume and $T_{el}=T_{ion}$, the free energy is obtained using Eqs. (4a-e). The free energy is then interpolated in $V$ using a 3$^{rd}$ order finite strain (Birch-Murnaghan) equation of state (EoS). When using the TQHA scheme, i.e., Eqs. (2a-h), the free energy is computed directly on the same $T_{el}$, $T_{ion}$, and $V$ grids and then interpolated for $T_{el}$ and $V$ as done in the EIM scheme. We also use the TIP scheme, i.e., the EIM approach with *T-independent* phonon frequencies. For comparison, we also compute free energies simply using the QHA (Eq. 1a-c) scheme with $T_{el} = $ cte. These four schemes - EIM, TQHA, TIP and QHA - help to elucidate the impact of $T_{el}$ and $T_{ion}$ on vibrational and thermodynamic properties. They are summarized in Table I. We compute thermodynamic properties using some modules of the `qha` Python package [40] after obtaining free energies with these four schemes.

## III. RESULTS AND DISCUSSIONS
### 1. Helmholtz free energy and entropy

As indicated in Section II, the PGM framework addresses four distinct situations: i) harmonic insulators and ii) metals and iii) anharmonic insulators, and iv) anharmonic metals. Except for i), all other cases have $T$-dependent frequencies. Free energy computations using the EIM approach is appropriate for all cases, but, in principle, it is necessary for cases ii) to iv). Here we also explore the performance of the TQHA approach, i.e., Eqs. (2).

Fig. 1 shows the vibrational density of state (VDoS) of ε-Fe for different $T_{el}$'s. Each VDoS is calculated for structures optimized at 360 GPa (static pressure), which vary slightly with $T_{el}$. These VDoS's do not differ drastically, suggesting a potentially weak dependence of the vibrational free energy on $T_{el}$. Fig. 2 shows the entropy and free energy of ε-Fe produced by the EIM and TQHA approaches. $S_{el}(T)$ and $F_{st}(T)$ (Eqs. (2e) and (2c)) are computed the same way in both approaches, but results differ in the way $S_{vib}(T)$ and $F_{vib}(T)$ are computed, i.e., using Eqs. (2g) and (2f) (TQHA) or Eqs. (4c) and (4b) (EIM). The negative $T$-derivative of $F_{vib}^{TQHA}(T)$ gives $S_{vib}^{TQHA}(T)$ (Eqs. 2f and 2g) while the negative integration of $S_{vib}^{EIM}(T)$ gives $F_{vib}^{EIM}(T)$ (Eq. 4c and 4b).

The dotted red lines in Figs. 2a and 2b display $S_{el}(T)$ (Eq. 2e) and $F_{st}(T)$ (Eq. 2c), respectively. Dashed colored lines display $S_{vib}^{T_{el}=cte}(T)$ (Eqs. 2g,h) and $F_{vib}^{T_{el}=cte}(T)$ (Eq. 2f-h), the latter being shifted by a constant, $-F_0$, the total static energy at $T_{el} = 0$. Colored circles are $S_{tot}^{TQHA}(T) = S_{el}(T) + S_{vib}^{T_{el}=T}(T)$ (Eqs. 2e,g) and $F_{tot}^{TQHA} = F_{st}(T) + F_{vib}^{T_{el}=T}(T) - F_0$ (Eqs. 2a-f). Solid black lines are $S_{tot}^{EIM}(T) = S_{el}(T) + S_{vib}^{EIM}(T)$ (Eqs. 2e, 4c, 4e) and $F_{tot}^{EIM} = F_{st}(T) + F_{vib}^{EIM}(T) - F_0$ (Eqs. 4a,b). As can be seen, $S_{vib}^{T_{el}=cte}(T)$ depends very weakly on $T_{el}$, while $F_{vib}^{T_{el}=cte}(T)$ is slightly more sensitive to it. Nevertheless, $S_{tot}^{EIM}(T) \cong S_{tot}^{TQHA}(T)$, indicating that the last term on the r.h.s. of Eq. (2g) is relatively insignificant in the case of ε-Fe and likely other metals as well. Also, $F_{tot}^{EIM}(T) \cong F_{tot}^{TQHA}(T)$ for ε-Fe. The similarity of these quantities appears to justify the use of a single VDoS in calculations of $F_{vib}(V,T)$ or $S_{vib}(V,T)$ for ε-Fe (TIP scheme).

## 2. Vibrational properties

The vibrational entropy, $S_{vib}$, needs special attention since any uncertainty in $S_{vib}$ propagates to other properties via the entropy integration in the EIM approach. Fig. 3 shows a comparison between $S_{vib}^{EIM}(P, 300\ K)$ and several experimental data sets of this quantity. All $S_{vib}^{exp}(P, 300\ K)$ shown in Fig. 3 were estimated using VDoS's obtained by nuclear resonant inelastic x-ray scattering (NRIXS) [41–45] and Eq. (1d) or (4c). The electronic entropy, $S_{el}$, does not contribute explicitly to $S_{vib}$.

A rapid decrease in $S_{vib}^{exp}$ around 13 GPa is generally attributed to the phase transition between α-Fe (bcc) and ε-Fe (hcp) [13,46]. $S_{vib}^{EIM}(P, 300\ K)$ agrees very well with all the reported NRIXS data, especially with the more recent one by Murphy et al [45] at higher pressures. At lower pressures (P ≤ 50 GPa), $S_{vib}^{EIM}(P, 300\ K)$ deviates slightly from $S_{vib}^{exp}(P, 300\ K)$. This is likely related to errors in the calculation of the exchange-correlation energy. As shown in the following section (Fig. 7), the 300 K compression curve of ε-Fe is not well reproduced computationally at low pressures. Despite this issue, our vibrational entropy agrees well with available experimental data which also show considerable uncertainty.

The vibrational pressure, $P_{vib}(V,T)$, calculated as $-\left(\frac{\partial F_{vib}}{\partial V}\right)_T$, captures more clearly the $T_{el}$-dependence of the VDoS. Fig 4. compares $P_{vib}^{EIM}(V, T = cte)$ and $P_{vib}^{TQHA}(V, T = cte)$ with $P_{vib}^{exp}(V, T = cte)$. The latter was estimated first using VDoS's measured with NRIXS [47], followed by calculations of $F_{vib}^{exp}$ and $-\left(\frac{\partial F_{vib}}{\partial V}\right)_T$, all at 300 K. This is the same procedure used in the calculation of $P_{vib}^{EIM}(V, T = cte.)$. The reported $P_{vib}^{exp}(V,T)$ [47] was modeled using the measured $P_{vib}^{exp}(V, 300\ K)$ and first extending it linearly to high $T$ and reported as the harmonic part, $P_{vib}^{h}(V,T)$. The anharmonic contribution was modeled [13] using *ab initio* anharmonic free energy calculations [48] with $P_{vib}^{anh}(V,T)$ fit to a phenomenological formulation [49]. Therefore, $P_{vib}^{exp}(V,T) = P_{vib}^{h}(V,T) + P_{vib}^{anh}(V,T)$. Here we see EIM and TQHA results are almost indistinguishable, meaning, the $T_{el}$-dependence of phonon frequencies is small and could be disregarded, as will be done in Figs. 5 and 6 by comparing EIM and TIP results. $P_{vib}^{exp}$ agrees equally well with $P_{vib}^{EIM}$ and $P_{vib}^{TQHA}$ up to 4000 K suggesting the anharmonic contribution, disregarded in our calculations, is negligible. At 5600 K we notice a deviation from $P_{vib}^{exp}$, which suggests anharmonic effects are important at this temperature and beyond. Fig. 4 also shows QHA results, meaning, $F_{st}(V,T)$ and $F_{vib}(V,T)$, are both computed using $T_{el}$ = cte, as indicated. This type of calculation is not uncommon for practical comparisons with experimental data at specific temperatures. At 5600 K, QHA results are not available since we do not have phonons with $T_{el}$ = 5600 $K$ but we show QHA results at 5000 K and 6000 K for comparison. The discrepancy between QHA and EIM/TQHA results indicates the importance of properly computing the static free energy, $F_{st}(V,T_{el})$, in a continuum (fine grid) of $T_{el} = T_{ion}$. As pointed out above, the agreement between EIM and TQHA results indicates that the effect $T_{el}$ on $F_{vib}(T,V)$ is negligible in ε-Fe. The origin of the deviation of QHA results from EIM/TQHA is clarified in the next calculation of thermal expansivity at 300 K.

### 3. Thermal expansion coefficient

The thermal expansion coefficient, $\alpha = \frac{1}{V}\left(\frac{\partial V}{\partial T}\right)_P$, is very sensitive to anharmonicity [22]. Discrepancies between *ab initio* results using the QHA and experimental data at high temperatures are often attributed to anharmonicity. However, the influence of thermal electronic excitations on this quantity in metallic systems is rarely addressed. To address this influence, we examine this property at 300 K where anharmonicity is expected to be insignificant.

As shown in Figs. 2 and 4, TQHA and EIM approach give quite similar values for entropy, free energy, and pressure for ε-Fe up to ~4,000 K. Therefore, we expect them to offer similarly good results for other thermodynamic properties as well. Here we examine the difference between $\alpha^{EIM}(P, 300\text{ K})$, $\alpha^{TIP}(P, 300\text{ K})$, and $\alpha^{QHA}(P, 300\text{ K})$. $\alpha^{TIP}(P, 300\text{ K})$ is computed using $T$-independent phonons obtained with $T_{el} = 300\ K$ and $\alpha^{QHA}(P, 300\text{ K})$ used $T_{el} = 300\ K$ to compute $F_{st}(V,T)$ also. The last procedure is standard and is inspired by the small $T_{el}$-dependence of the VDoS. Fig. 5 shows these quantities and compares them with $\alpha^{exp}(P, 300\text{ K})$ [50]. First, at these relatively low temperatures, the $T_{el}$-dependence of the VDoSs has little impact on this quantity, with $\alpha^{EIM}$ and $\alpha^{TIP}$ results being quite similar. Second, the $T_{el}$-dependence of $F_{st}(V,T)$, not included in $\alpha^{QHA}$, strongly affects this quantity. Because the calculation of $\alpha$ involves a $T$- derivative, the influence of $T_{el}$ on $F_{st}(V,T)$ and consequently on $P_{st}(V,T)$, or $V(P,T)$, considerably affects $\alpha(P,T)$ even at low temperatures, as previously pointed out [30].

The inset in Fig. 5 compares $\alpha^{EIM}(P,T)$ and $\alpha^{TIP}(P,T)$ with $\alpha^{exp}(P,T)$ at much higher pressures and temperatures. $\alpha^{exp}(P,T)$, was obtained using a semi-empirical EoS for ε-Fe [13], which was computed based on DAC experiments at ambient temperature, Hugoniot data [51], and some *ab initio* modeling [48]. *Ab initio* results by Sha & Cohen [4] are also presented for comparison. First, $\alpha^{EIM}(P,T)$ agrees with $\alpha^{exp}(P,T)$ the best. Second, $\alpha^{EIM}(P,T)$ and $\alpha^{TIP}(P,T)$ increasingly deviate from each other with increasing temperatures but the deviation decreases with increasing pressure. This indicates that the $T$-dependence of phonon frequencies, whether originating in electronic excitations of phonon-phonon interactions becomes relevant at these high temperatures. Since the discrepancy between $\alpha^{EIM}(P,T)$ and $\alpha^{exp}(P,T)$ decreases with increasing pressure, anharmonicity should be the source of this discrepancy, as indicated by $P_{vib}$ calculations (Fig. 4). Further improvement in the agreement between these quantities is expected by replacing phonon frequencies with quasiparticle frequencies [21,52] more properly describing the $T$-dependence of the VDoS.

## 4. Other Thermodynamic Properties

With previously obtained quantities, we compute the internal energy $E = F + TS$ and other thermodynamic properties such as the isothermal bulk modulus, $K_T = -V\left(\frac{\partial P}{\partial V}\right)_T$, and isochoric and isobaric specific heat, i.e., $C_V = \left(\frac{\partial E}{\partial T}\right)_V$ and $C_P = C_V + \alpha^2 TVK_T$, as shown in Fig. 6. We compare semi-empirically "modeled" experimental data [13] with results of the EIM and TIP schemes, the latter being common and much more computationally efficient.

As shown in Fig. (6a), $K_T^{EIM}(P,T)$ results agree well with $K_T^{exp}(P,T)$ as do results from a previous calculation [22]. Differences depend slightly on temperature and pressure. It increases with increasing temperature and decreasing with pressure. This behavior is usually the symptom of anharmonicity. Second, the deviation between $K_T^{EIM}(P,T)$ and $K_T^{TIP}(P,T)$ is quite visible above 4,000 K at all pressures, while the nature of this deviation changes with increasing pressure. This implies the $T_{el}$-dependence of the VDOs is impactful. Third, $K_T^{EIM}(P,T)$ also agrees very well with *ab initio* molecular dynamics (MD) results [22], $K_T^{MD}(P,T)$, that account for anharmonic effects up to 6000 K and 400 GPa. At higher pressures, a noticeable deviation develops between $K_T^{EIM}(P,T)$ and $K_T^{MD}(P,T)$. The origin of this discrepancy is unclear, but it could be technical issues such as the choice of pseudopotentials.

$C_V(P,T)$ and $C_P(P,T)$ shown in Figs. 6(b,c) indicate very similar trends: $C_{V,P}^{EIM}(P,T)$ agree better with modeled experimental data [13] than $C_{V,P}^{TIP}(P,T)$. The difference between $C_{V,P}^{EIM}(P,T)$ and $C_{V,P}^{exp}(P,T)$ increases at higher temperatures and decreases at higher pressures. Similarly, the difference between $C_{V,P}^{EIM}(P,T)$ and $C_{V,P}^{TIP}(P,T)$ increases at higher temperatures and decreases at higher pressures. As expected, the $T$-dependence of $C_{V,P}^{EIM}(P,T)$ is nearly linear at high temperatures, but we notice a deviation from linearity around 4,000 K at 200 GPa, which decreases at higher pressures. Such deviation reflects the sampling of the non-parabolic electronic density of states of $\varepsilon$-Fe around the Fermi level at these temperatures.

In summary, above 4,000 K we see indications of anharmonic effects on thermodynamic properties at pressures above 200 GPa. $C_V$ and $C_P$ suggest anharmonic effects decrease with increasing pressure.

## 5. Equations of state at 300 K

Fig. 7 shows the calculated compression curve of $\varepsilon$-Fe compared with other *ab initio* calculations [2,4,9,53] and DAC experimental data [11–13,54–56] at 300 K. $V^{EIM}(P, 300\,K)$ agrees well with most *ab initio* predictions, especially at the highest pressures, except with Sha & Cohen's results [4]. All spin-unpolarized calculations using the PBE-GGA deviate from $V^{exp}(P, 300\,K)$ curves below ~ 100 GPa. Above ~100 GPa, which is the pressure range of concern here, essentially all *ab initio* predictions show very good agreement with experimental data. There is also some deviation between various $V^{exp}(P, 300\,K)$ data sets, especially at the highest pressures. Thus, *ab initio* predictions are highly reproducible and reliable at such high pressures and should provide good constraints on the properties of $\varepsilon$-Fe. Combining this conclusion with conclusions from previous sections, it seems that as long as $T_{el}$ effects are properly included also in VDoS calculations, the EIM approach can offer reliable properties of $\varepsilon$-Fe without including anharmonic effects up to 360 GPa and ~4,000 K, somewhat shy from inner core temperatures. Reaching inner core temperatures requires inclusion of anharmonic effects as done in other *ab initio* calculations (e.g., [2,20,22]).

## 6. Isentropic equations of state (EoS)

The following results are obtained using the EIM scheme. To generate an isentropic EoS, we first compute the adiabatic gradient as follows,

$$\left(\frac{\partial T}{\partial P}\right)_S = \frac{\alpha V T}{C_P}, \tag{5a}$$

where $\alpha$ is the thermal expansion coefficient and $C_P$ is the isobaric specific heat. The adiabatic temperature profile is then obtained by integrating the adiabatic gradient with appropriate initial conditions:

$$T = T_0 + \int_{P_0}^{P_1} \left(\frac{\partial T}{\partial P}\right)_S dP. \tag{5b}$$

We assume the initial conditions, $P_0 = 60$ GPa and $T_0 = 1050$ K, to reproduce the experimental ramp compression data on Fe obtained at the National Ignition Facility (NIF) [16]. We then compute the volume/density throughout this $T - P$ profile. We use three different valence-core electron interaction potentials, EPAW [32], JTH [33] and GBRV [34] to assess the impact of this approximation at very high pressures, a significant point of concern. The vibrational contribution to the free energy is computed using VDoSs calculated using the EPAW method. The impact of pseudopotential changes on $F_{vib}$ is of second order compared to the impact on $F_{st}$ [57].

The possible $T - P$ profiles reported in the NIF experiment and our calculated profiles are shown in the Fig. 8 inset. The predicted isentropes are well within the range of temperatures expected in the NIF experiments. Our predicted isentropic $P - \rho$ EoSs using all three potentials also agree reasonably well with the NIF data up to 1,000 GPa (Fig. 8(b)).

The JTH potential results agree best with the experimental data, but it does not reproduce the all-electron static EoS. The GBRV and EPAW predictions agree well with each other. EPAW reproduces better the all-electron EoS [32]. This seems to suggest that the unsatisfactory performance of these good EPAW and GBRV potentials is related to the uncertain temperature in the NIF experiments at the highest pressure. The experimental compression curve may not follow precisely an isentrope, and the temperature achieved in the experiments might be slightly higher than the temperature range represented in the Fig. 8 inset. On the other hand, we have not included anharmonic effects in our calculations, even though the isentropic temperature profile does not seem to reach temperatures at which this issue should be of concern. Nevertheless, anharmonicity should be included for greater accuracy in such calculations and for a more conclusive diagnostic on the origin of the discrepancy between the measured and calculated isentropic $P - \rho$ profile.

### 7. Equations of state on the Hugoniot

Shockwave data gives Hugoniot EoSs in a wide pressure-temperature range, i.e., at least up to 250 GPa and 5,000 K for $\varepsilon$-Fe. Conservation of mass, momentum, and energy in experiments produce the Rankine-Hugoniot formula [1,58]

$$\frac{1}{2} P_H (V_0 - V_H) = E_H - E_0, \tag{6}$$

where $P$ is pressure, $V$ is volume, $E$ is internal energy. The subscript $H$ on $V$, $P$ and $E$ stands for Hugoniot. We calculate the $T - P$ path such that the Hugoniot relation in Eq. (6) is observed. Our predicted EIM EoS along the Hugoniot and the associated $T - P$ path are shown in Fig. 9. Our EIM predictions agree well with experiments [14,51,59] and other *ab initio* calculations [4,48] up to ~ 4000 K and ~ 200 GPa.

### IV. CONCLUSION

We have implemented an accurate free energy calculation scheme based on the entropy integration that applies equally to situations where phonon frequencies are $T$-dependent or -independent. Several free energy computation schemes have been compared, showing that the entropy integration method is the most accurate, especially when dealing with metallic systems. Using $\varepsilon$-Fe as a test case, we have explored the detailed performance of this method to compute several thermodynamic properties. For metallic systems, the effect of thermal electronic excitations on the vibrational free energy is not significant at up to ~4,000 K at ~200 GPa in the present case, which justifies the use of $T$-independent frequencies in many practical situations. However, accurate predictions of metals' thermodynamic properties require computations of the Mermin free energy, the static energy component, in a $T_{el}$-continuum. This effect is clearly demonstrated in the calculation of the thermal expansion coefficient of $\varepsilon$-Fe at 300 K. Anharmonic effects do require consideration of $T$-dependent frequencies (quasi-particle frequencies) to be predictive.

The computed vibrational entropy, vibrational pressure, thermal expansion coefficient, and equation of state are in overall good agreement with measurements at least up to 4,000 K at 200

GPa when thermal electronic excitations are properly accounted for. Beyond these conditions, anharmonicity needs to be addressed in these calculations, which can be accomplished by replacing phonon frequencies with $T$-dependent phonon quasiparticle frequencies [24]. We expect this procedure to be predictive in computing properties of hot iron cores in exoplanets.

***Acknowledgments*** - This work was supported in part by the US Department of Energy award DESC0019759 and in part by the National Science Foundation award EAR-1918126 (R.M.W.). This work used the Extreme Science and Engineering Discovery Environment (XSEDE), USA, which was supported by the National Science Foundation, USA Grant Number ACI-1548562. Computations were performed on Stampede2, the flagship supercomputer at the Texas Advanced Computing Center (TACC), The University of Texas at Austin generously funded by the National Science Foundation (NSF) award ACI-1134872.

**Table and Figures**

| Scheme | Name | Free energy calculations | | | Validity |
|---|---|---|---|---|---|
| | | Formalism | $T_{el}$ effect on $F_{st}$ | $T_{el}$ effect on $F_{vib}$ | |
| EIM | Entropy integration method | Eqs. (4a-e) | Yes | Yes | Anharmonic and nearly harmonic metals and insulators |
| TQHA | Temperature-dependent quasi-harmonic approximation | Eqs. (2a-h) | Yes | Yes | Good approximation for nearly harmonic metals and insulators |
| TIP | Temperature-independent phonons | Eqs. (4a-e) | Yes | No | Good approximation for nearly harmonic metals and insulators |
| QHA | Quasi-harmonic approximation | Eqs. (1a-c) | No | No | Nearly harmonic insulators |

**Table I** - Free energy computation schemes used in this study.

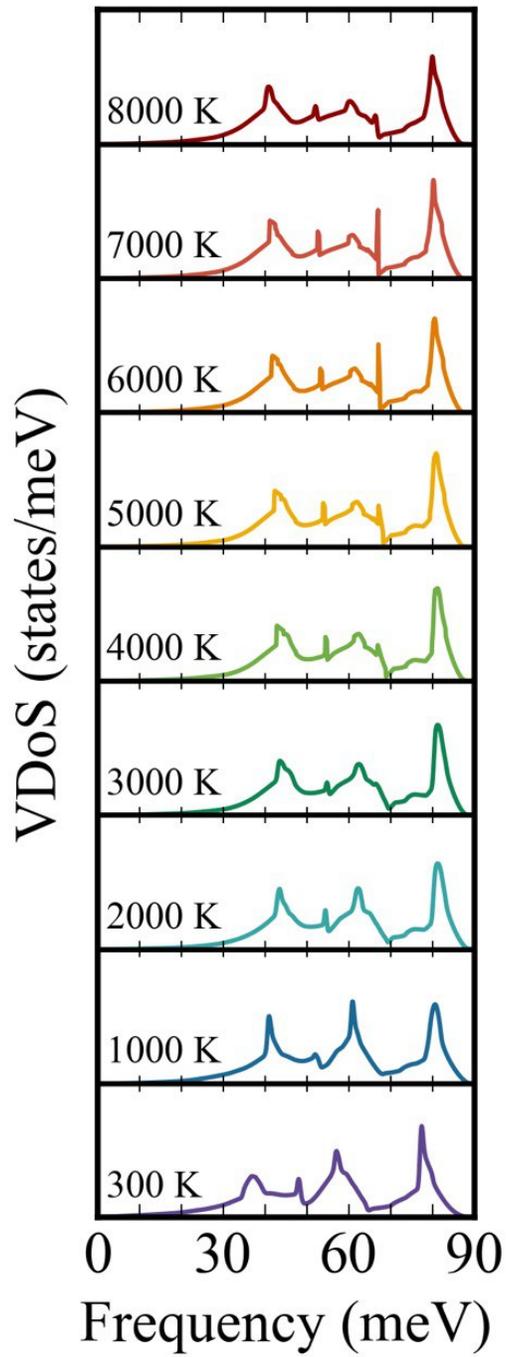

**Figure 1** - Vibrational density of states of hcp iron optimized at various electronic temperatures, $T_{el}$, at 360 GPa.

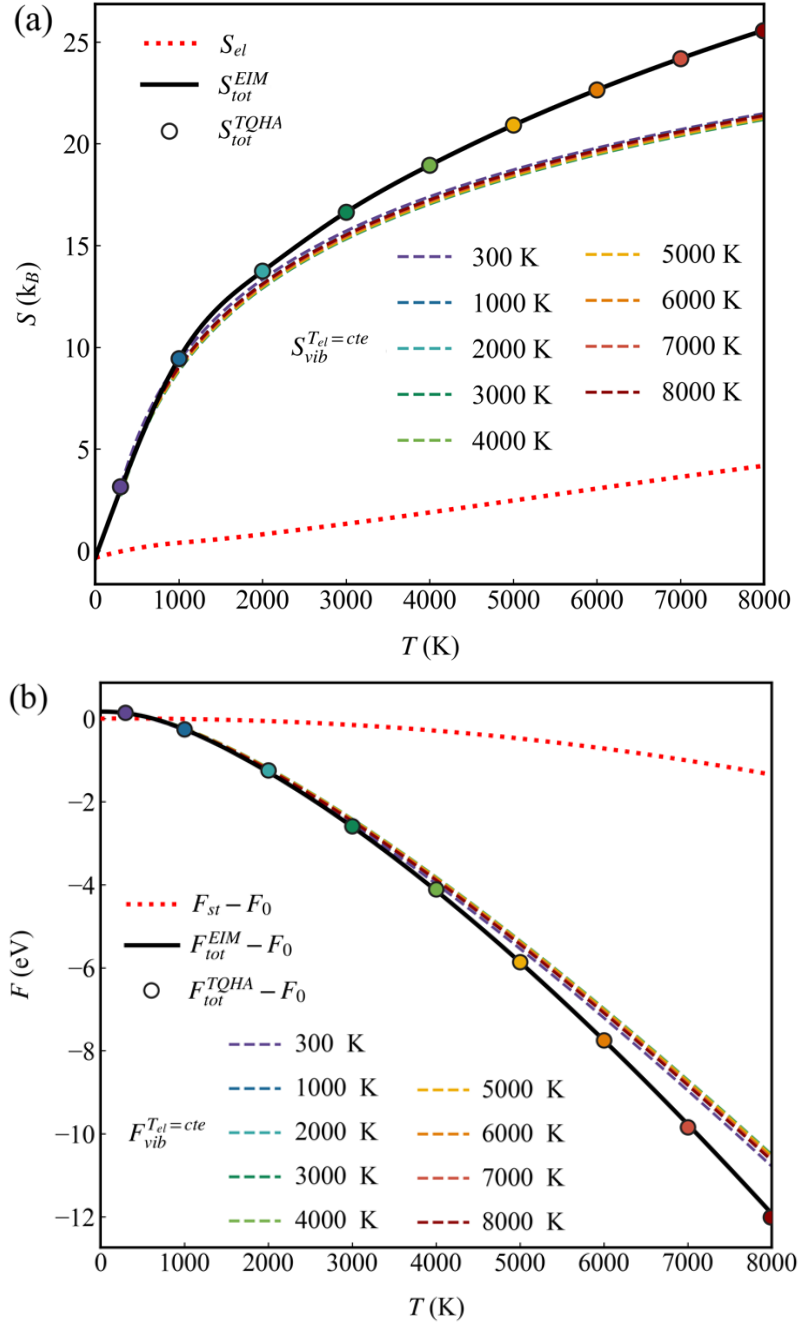

**Figure 2** - (a) Entropy at constant volume ($V = 6.77$ Å³/atom). Dotted red line: $S_{el}(T_{el} = T)$ (Eqs. (2d,e)). Dashed colored lines: $S_{vib}^{T_{el}=cte}(T_{ion} = T)$ (Eq. (2g)). Colored circles: $S_{tot}^{TQHA}(T) = S_{el}(T_{el} = T) + S_{vib}^{T_{el}=T}(T_{ion} = T)$. Solid black line: $S_{tot}^{EIM}(T_{el} = T_{ion} = T)$ (Eqs. (4c,d)). (b) Free energy at the same $V$. Dotted red line: $F_{st}(T_{el} = T) - F_0$ (Eq. 2c), where $F_0 = -8913.5$ eV is the static energy at $T = 0$ K. Dashed colored lines: $F_{vib}^{T_{el}=cte}(T_{ion} = T)$ (Eqs. 2a,b,f) at various $T_{el}$'s. Colored circles: $F_{tot}^{TQHA} = F_{st}(T_{el} = T) + F_{vib}^{T_{el}=T}(T_{ion} = T) - F_0$ (Eqs. 2a-h). Solid black line: $F_{tot}^{EIM} = F_{st}(T_{el} = T) + F_{vib}(T_{el} = T_{ion} = T) - F_0$ (Eqs. 4a,b).

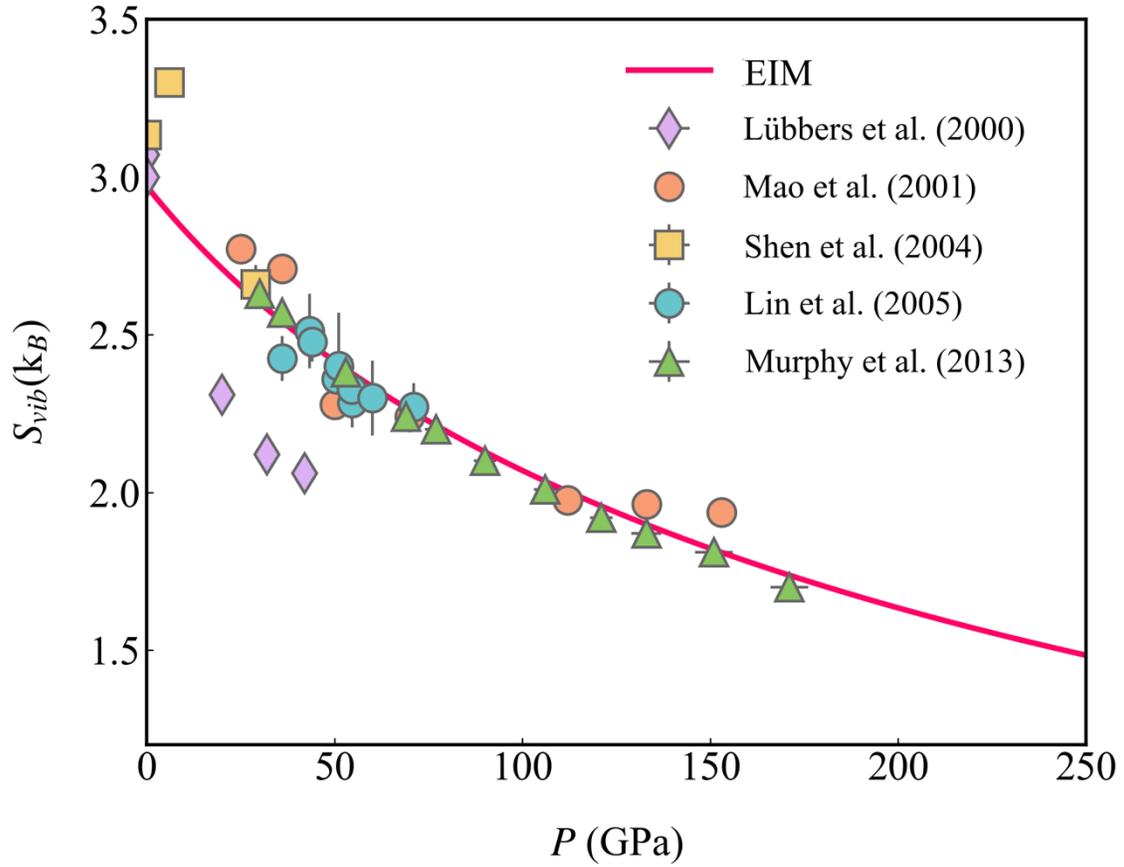

**Figure 3** - Vibrational entropy of ε-Fe at $T_{el} = T_{ion} = 300\ K$. The solid line is $S_{vib}^{EIM}(P, T = 300\ K)$ (Eq. (4c)) and symbols are experimental values, $S_{vib}^{exp}(P, 300\ K)$, obtained using Eq. (1d) and VDoS's measured by NRIXS [41–45].

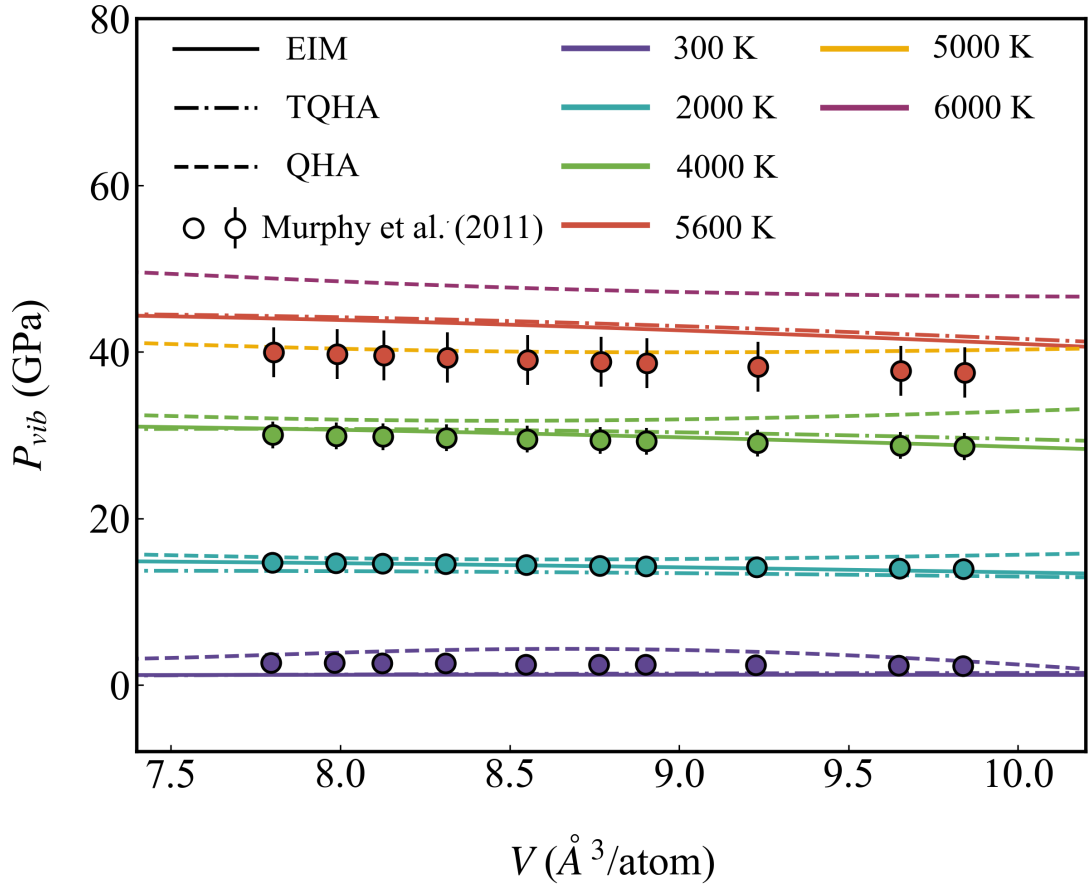

**Figure 4** - Vibrational pressure of ε-Fe. Solid lines are $P_{vib}^{EIM}(V,T)$; dashed lines are QHA calculations at $T_{el} = cte$ (see text); dash-dotted lines are $P_{vib}^{TQHA}(V,T)$. QHA results are not available at 5600 K because we did not generate VDoSs at $T_{el}$= 5600 K, but they are bracketed by 5000 K and 6000 K values. Symbols are $P_{vib}^{exp}(V,T)$ obtained from VDoS's measured using NRIXS [47] at 300 K and modeled harmonic and anharmonic contributions at higher $T$s. See text.

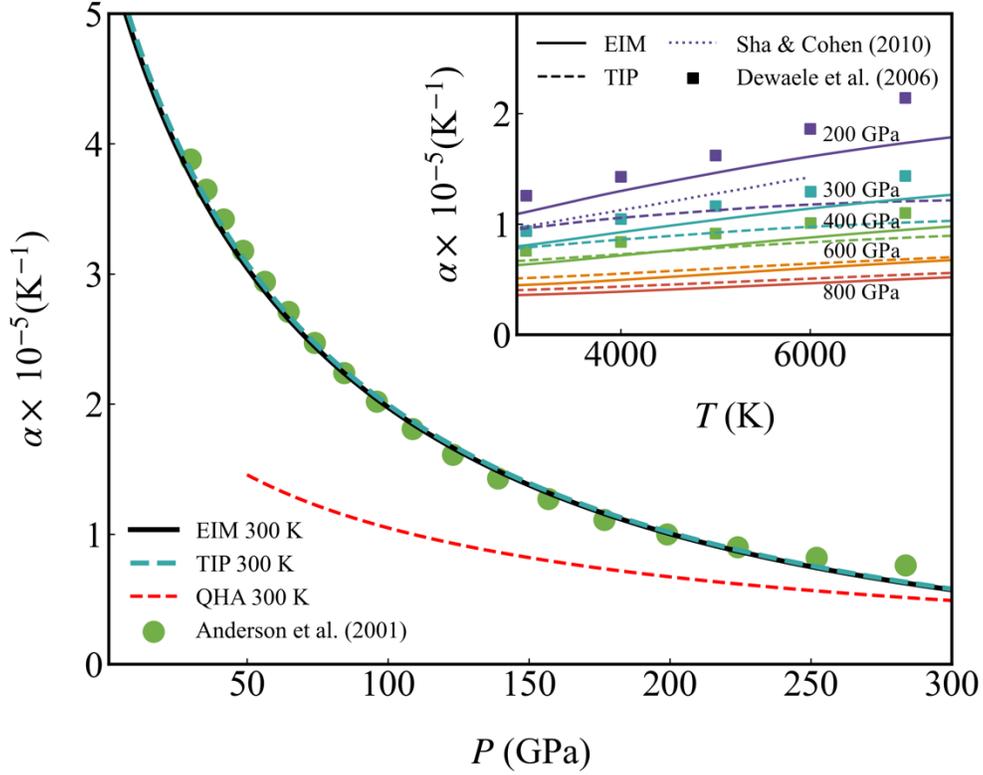

**Figure 5** - Thermal expansion coefficient of ε-Fe at 300 K. $\alpha^{EIM}(P, 300\ K)$ (solid black line) and $\alpha^{TIP}(P, 300\ K)$ (dashed blue line) are quite similar, indicating the low impact of a variable $T_{el}$ on the vibrational free energy, $F_{vib}$. The difference between $\alpha^{EIM}(P, 300\ K)$ and $\alpha^{QHA}(P, 300\ K)$ (dashed red line) reveals the significant impact of the variable $T_{el}$ in the static free energy, $F_{st}$, computed using the Mermin functional. As expected, $\alpha^{EIM}(P, 300\ K)$ agrees best with $\alpha^{exp}(P, 300\ K)$ (green circles) [50]. *Inset* - solid lines are $\alpha^{EIM}(P = cte, T)$ and dashed lines are $\alpha^{TIP}(P = cte, T)$ obtained using the VDoSs computed with $T_{el} = 300\ K$. Dotted lines are *ab initio* results by Sha and Cohen [4]. Squares are semi-empirically modeled [60] experimental data obtained in DAC experiments [13].

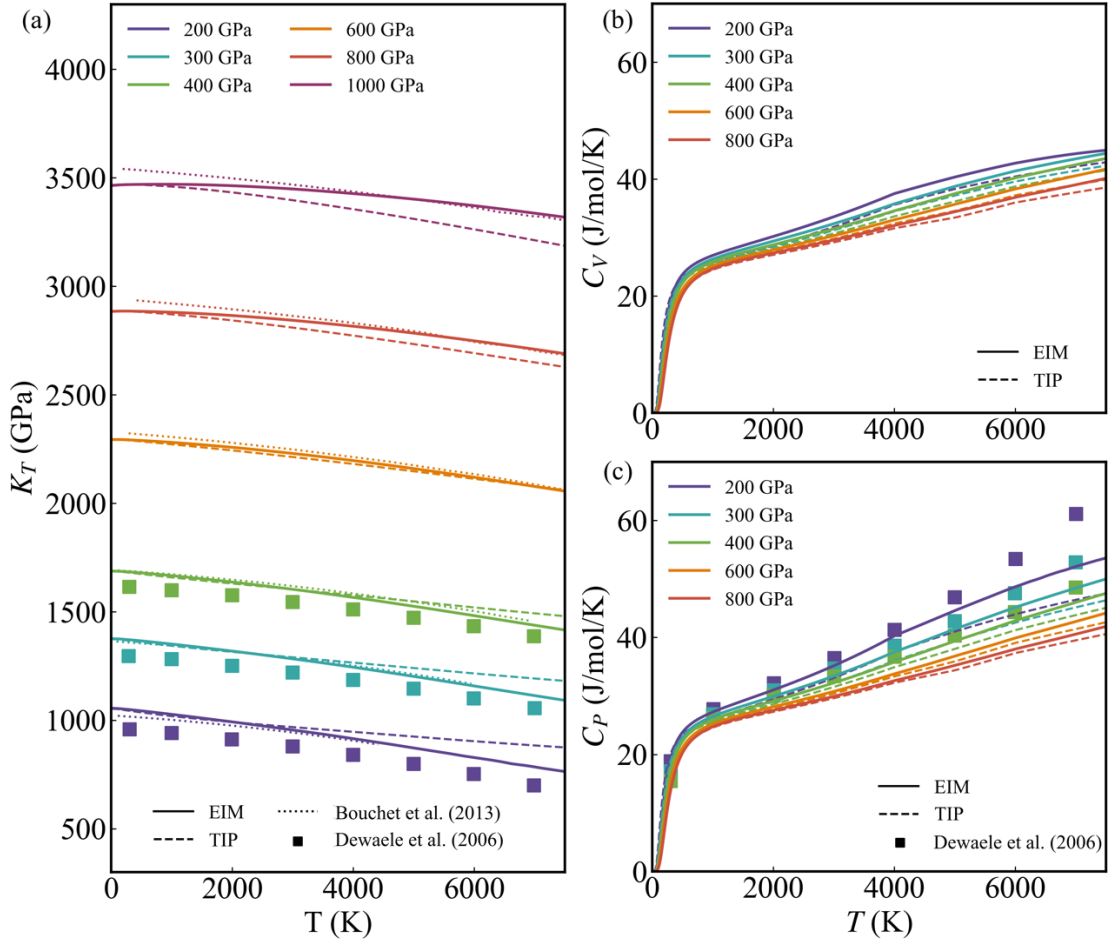

**Figure 6** - (a) Isothermal bulk modulus, $K_T$, (b) isochoric specific heat, $C_V$, and (c) isobaric specific heat, $C_P$, of ε-Fe. Solid lines are obtained using the EIM method; dashed lines are obtained using the TIP scheme with phonons calculated with $T_{el} = 300$ K. Dotted lines in a) are from a previous *ab initio* study [22]. Symbols are results of a thermodynamic analysis [60] based on equations of state measured in DAC experiments [13].

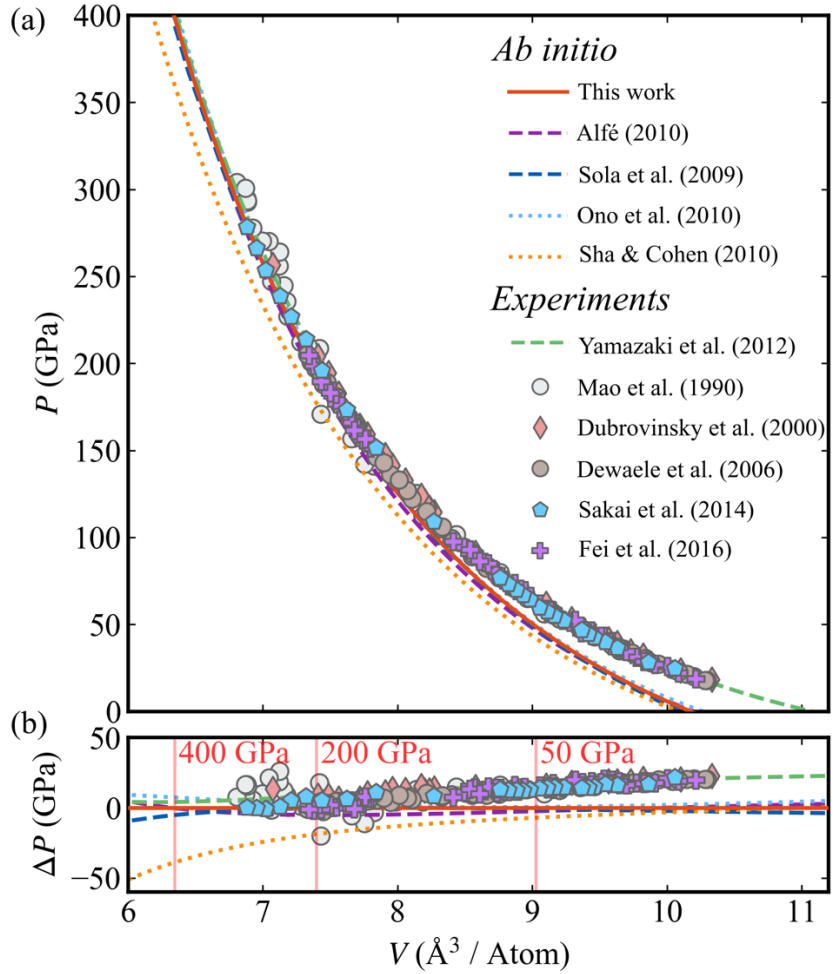

**Figure 7** - (a) Comparison between experimental and theoretical equations of state (EoS) of ε-Fe at 300 K. Symbols and dashed green line were obtained in static compression experiments on a diamond anvil cell (DAC): Mao et al. [11], Dubrovinsky et al. [12], Dewaele et al. [13], Sakai et al. [54], Fei et al. [55], and Yamazaki et al. [56]. The latter was reported as parameters of a Birch–Murnaghan 3$^{rd}$ order (BM3) [61] equation of state. Theoretical results include *ab initio* PBE-MD results by Alfé [9], quantum Monte Carlo (QMC) by Sola et al. [2], static PBE results by Ono et al. [53], PBE-LMTO results by Sha and Cohen [4]. Results reported as equation of state parameters (BM3 and Vinet [62]) EoS parameters are shown as dashed and dotted lines respectively. (b) Pressure difference vs. $V$, $\Delta P(V)$, w.r.t. our EIM results. Vertical lines at $P = 50$ GPa, 200 GPa, 400 GPa indicate volumes used in our EIM calculations.

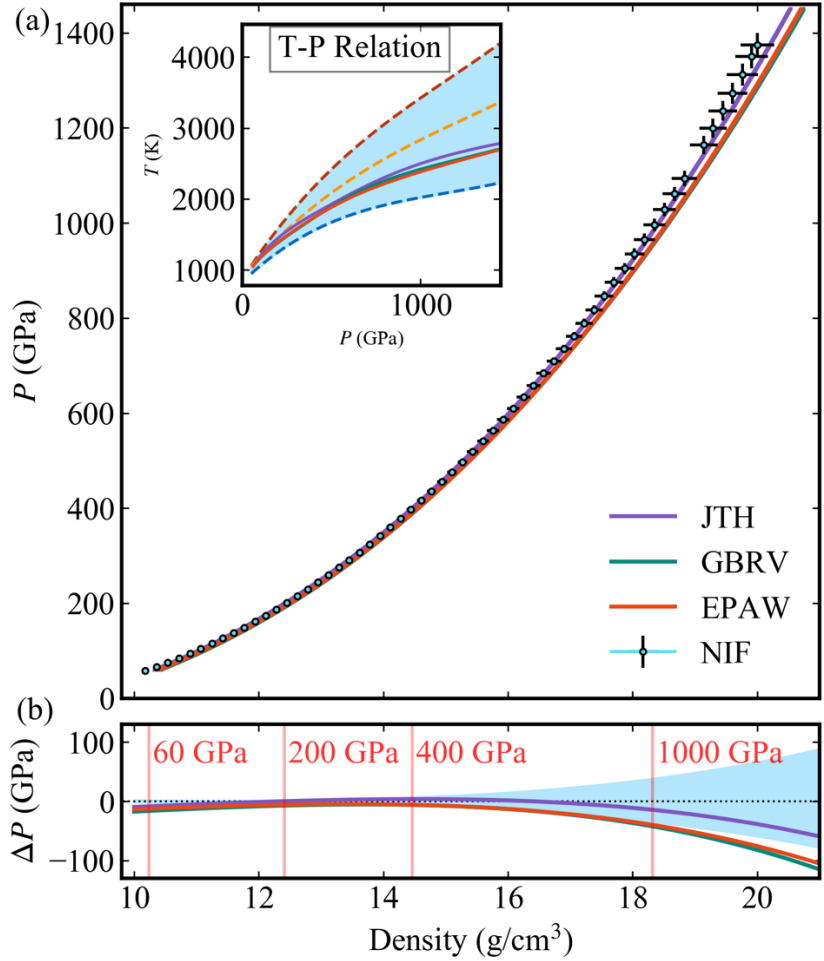

**Figure 8** - (a) Isentropic $P - \rho$ EoS of ε-Fe. Small circles with uncertainties are the NIF data [16]. Solid lines correspond to EIM results and different pseudopotentials in the static part of the calculation: EPAW potential [32] (red line), JTH potential [33] (purple line), and GBRV ultra-soft potential [34] (green line). The vibrational contribution to the free energy in these calculations was obtained using the EPAW method. *Inset* - the temperature in the NIF data: red dashed line corresponds to an intermediate strength model for Fe; yellow dashed line is a low-strength model; blue dashed line corresponds to shock compression to 60 GPa before ramp compression to pressures over 1 TPa; blue shaded region corresponds to the possible temperature range achieved in the NIF experiment. Red, purple and green solid lines show adiabatic $T - P$ relations obtained using the EIM method and EPAW, JTH, and GBRV potentials, respectively. (b) Pressure difference w.r.t. to the NIF data (circles with uncertainties), $\Delta P(\rho)$: blue shaded region shows possible uncertainty in the NIF data; red, purple, and green lines correspond to our EIM results using different potentials in the static part of the calculations along their respective adiabats shown in the inset in (a). Red vertical lines show corresponding pressures in the NIF data.

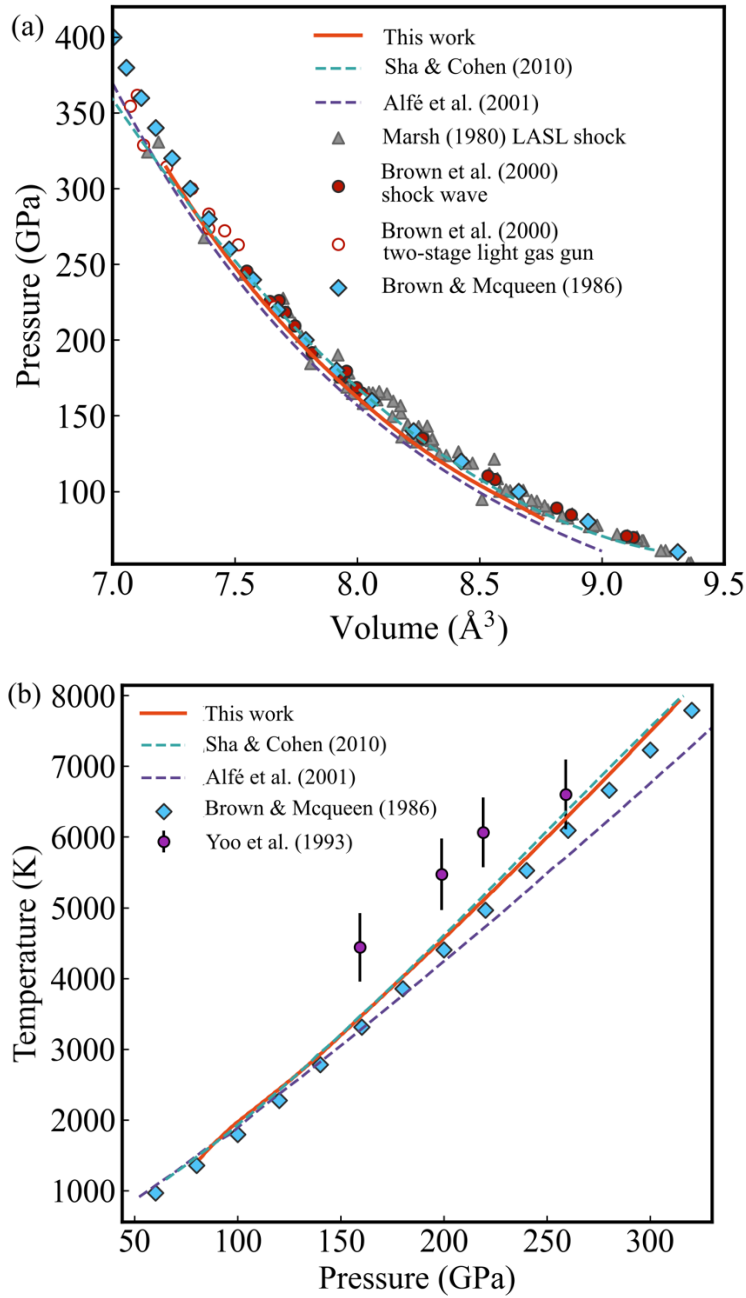

**Figure 9** - (a) *P* vs. *V* and (b) *T* vs. *P* along the Hugoniot. Solid red lines are EIM results using EPAW potentials; dashed blue lines are *ab initio* results by Sha and Cohen [4]; dashed purple lines are *ab initio* results by Alfé et al. [48]. Symbols correspond to experimental data: grey triangles from Ref. [59], filled red circles and the hollow circles from Ref. [51], filled blue diamonds from Ref. [14], purple circles with error bars from Ref. [15].

29